\documentclass[aps,twocolumn,superscriptaddress,nofootinbib,floatfix,a4paper,longbibliography]{revtex4-2}

\usepackage{times}
\usepackage{epsfig}
\usepackage{amsfonts}
\usepackage{amsthm}
\usepackage{amssymb}					
\usepackage{amsthm}
\usepackage{dsfont}
\usepackage{bm}
\usepackage{color}
\usepackage{multirow}
\usepackage[normalem]{ulem}
\newcommand{\stkout}[1]{\ifmmode\text{\sout{\ensuremath{#1}}}\else\sout{#1}\fi}
\usepackage{latexsym}
\usepackage{natbib}
\usepackage{verbatim}
\usepackage[T1]{fontenc}
\usepackage{float}
\usepackage{graphicx}
\usepackage{subcaption}
\usepackage{physics}
\usepackage{soul}
\usepackage{caption}
\usepackage{diagbox}
\usepackage{bbm}
\usepackage{tablefootnote}
\usepackage{flushend}
\usepackage{adjustbox}
\usepackage[dvipsnames]{xcolor}
\usepackage[font=small,labelfont=bf]{caption}

\usepackage{xspace}

\theoremstyle{definition}

\usepackage{amssymb}
\usepackage[customcolors]{hf-tikz} 
\usetikzlibrary{patterns}
\usetikzlibrary{matrix,decorations.pathreplacing}

\pgfkeys{tikz/mymatrixenv/.style={decoration={brace},every left delimiter/.style={xshift=8pt},every right delimiter/.style={xshift=-8pt}}}
\pgfkeys{tikz/mymatrix/.style={matrix of math nodes,nodes in empty cells,left delimiter={[},right delimiter={]},inner sep=1pt,outer sep=1.5pt,column sep=2pt,row sep=2pt,nodes={minimum width=20pt,minimum height=10pt,anchor=center,inner sep=0pt,outer sep=0pt}}}
\pgfkeys{tikz/mymatrixbrace/.style={decorate,thick}}

\usepackage[colorlinks=true,linkcolor=blue,citecolor=magenta,urlcolor=blue]{hyperref}

\begin{document}
	

\title{Entanglement in prepare-and-measure scenarios without receiver inputs}

\author{Elna Svegborn}
\affiliation{Physics Department and NanoLund, Lund University, Box 118, 22100 Lund, Sweden.}

\author{Armin Tavakoli}
\affiliation{Physics Department and NanoLund, Lund University, Box 118, 22100 Lund, Sweden.}
	
\begin{abstract}

The most elementary prepare-and-measure scenarios have no independent measurement inputs. No inputs mean that quantum advantages require two indispensable ingredients:  shared entanglement and measurements that can be adapted to the communicated messages. Understanding these scenarios is therefore  conceptually natural, but also practically relevant, since they act as testbeds for black-box certification of adaptive one-way LOCC. Here, we study them systematically and reveal several of their basic features. For classical messages, we first identify the minimal scenario with a quantum advantage and show that it is maximised by high-dimensional entanglement. Then, we identify the next-to-minimal scenario, and show that quantum advantages can be propelled by nonlocality of the Clauser-Horne-Shimony-Holt type, which makes this an appropriate setting for certification experiments. Proceeding further, we replace classical messages with quantum messages, but require  the receiver to read the message before measuring the entangled particle. We show that this leads to amplified quantum advantages, that are made possible only thanks to non-projective message read-out. This in dispensable role of  non-projective measurements challenges the common wisdom that they play a secondary role in revealing the power of quantum correlations in  black-box experiments.
\end{abstract}
	
\date{\today}

\maketitle
	
\textit{Introduction---}
Entanglement is a pivotal resource for communication. Although  it cannot by itself transmit information, shared entanglement can boost the  ability of  communicating parties to perform a distributed computation \cite{Cleve_1997,Brassard_2001}. These communication complexity advantages are fundamentally tied to Bell nonlocality \cite{Brukner_2004,Buhrman_2016,Tavakoli_2020}. These protocols typically proceed in two steps:  first the parties measure their share of the entangled state to obtain correlations that violate a Bell inequality, and then they use these correlations as advice for how to encode and decode  classical communication  \cite{Buhrman_2010}. However, it has been brought to attention that more general use of entanglement is possible, where one party can delay measuring one share of the entangled state until after the other party's message has arrived, and thereby adapting the measurement to the communication in real time \cite{Pauwels_2022}. These adaptive protocols, based on classical feed-forward, have been shown to  further boost quantum advantages.

The prepare-and-measure (PM) scenario, featuring a sender (Alice) and a receiver (Bob), is the standard setting to study these types of questions. In general, Alice and Bob both hold independent inputs and their job is to probabilistically compute functions of both inputs, while only being allowed limited communication resources. Given the ubiquity of the  PM scenario (see the review \cite{Brask_2026}), it is natural to take an interest in their simplest form. In the case, Bob has no independent input, and thus his goal is to learn some property about Alice's input. In addition to being the most elementary setting, the PM scenarios without measurement inputs are particularly interesting because --- in contrast to scenarios where Bob has inputs --- they do not admit any quantum advantage by replacing classical messages with quantum messages of the same dimension \cite{Frenkel_2015}. This means that quantum advantages are impossible unless the parties share entanglement. Moreover, when Bob has no input, the traditional approach to communication complexity, based on first violating a Bell inequality and then communicating classically, will always fail since Bell nonlocality is impossible without measurement inputs  \cite{Massar_2003}. This leads to a crucial observation: \textit{every} quantum advantage must rely on shared entanglement and measurements that adapt in real time to the communicated messages. 

In addition to theoretical interest, PM scenarios without independent measurement inputs also have a distinct applied motivation. Because the only quantum advantages are entanglement-based and of adaptive nature, such protocols can be interpreted as black-box certificates of adaptive one-way LOCC, i.e.~certifying that an experiment has the capacity to measure one part of a state and use the outcome in real time to inform what operations to perform on the other part of the state. Such feed-forward capability is broadly relevant in quantum technology, for instance in teleportation, quantum networks, quantum algorithms and quantum error correction.  The associated tests in the PM scenario are  device-independent, modulo the minimal assumption that the alphabet of the classical communication is restricted.


However,  beyond the  fact that they exist \cite{Frenkel_2022,Vieira_2023,Mir_2023,Rout_2025}, little is known about quantum advantages in PM scenarios without independent receiver inputs. Given their conceptual and practical interest, it is relevant to fill this gap of knowledge. In this work, we systematically reveal their key basic features. We focus on the elementary setting where the communication is one bit and identify the minimal input/output alphabets necessary to make a quantum advantage possible. We find that this minimal scenario admits a simple task-interpretation and 	derive the optimal protocol based on two-qubit entanglement, which turns out to use non-maximally entangled states. Furthermore, already in this minimial scenario, we find that  entanglement of higher-than-qubit dimension is necessary for the maximal quantum advantage. However, in order to pave the way for practical certification of one-way adaptive LOCC, one would ideally seek a test that (i) uses maximally entangled qubits, (ii) has good noise-resilience, and (iii) has input/output alphabets as small as possible. We meet all three desiderata by examining the next-to-minimal scenario, where we show that violations of the Clauser-Horne-Shimony-Holt (CHSH) Bell inequality can be linked to quantum advantages in the PM scenario. We then venture further and upgrade the communication from a bit to a qubit. In line with the adaptive nature of the protocol, we require Bob to first read the quantum message and then use the classical read-out to advice his choice of measurement on his entangled particle. We find that in spite of qubit messages alone not offering any advantage over bits \cite{Frenkel_2015}, using them as advice for which measurements to perform on entangled particles does lead to further advantages. Importantly, this general type of advantage is impossible unless one uses non-projective measurements, thereby giving them a distinct operational relevance in quantum correlations studies. 

\textit{Preliminaries.---} Consider the PM scenario where Alice selects an input, $x\in\{0,\ldots,|X|-1\}$, encodes it into a classical message $m\in\{1,\ldots,d\}$  and sends it to Bob who decodes it into an output $b\in\{0,\ldots,|B|-1\}$.  This leads to a probability distribution $p(b|x)$. Note that for this to be non-trivial we need $d<|X|$; otherwise Alice can just send her input Bob. In this classical approach, Alice and Bob may coordinate their encoding/decoding strategies with  a shared random variable, $\lambda$, subject to some distribution $p(\lambda)$. Hence, the correlations admit a  classical model if they can be written as $p(b|x) =   \sum_{\lambda,m} p(\lambda) p(m|x,\lambda) p(b|m,\lambda)$. The space of classical correlations is geometrically represented as a polytope. Therefore, it can be characterised by inequalities of the form \cite{Gallego_2010}
\begin{equation}\label{eq:facet}
\mathcal S = \sum_{b,x} c_{bx} p(b|x) \leq \beta,
\end{equation}
where $c_{bx}$ are some real-valued coefficients and $\beta$ is a tight bound satisfied by all classical models.

\begin{figure}[t!]
	\centering
	\includegraphics[width=0.8\columnwidth]{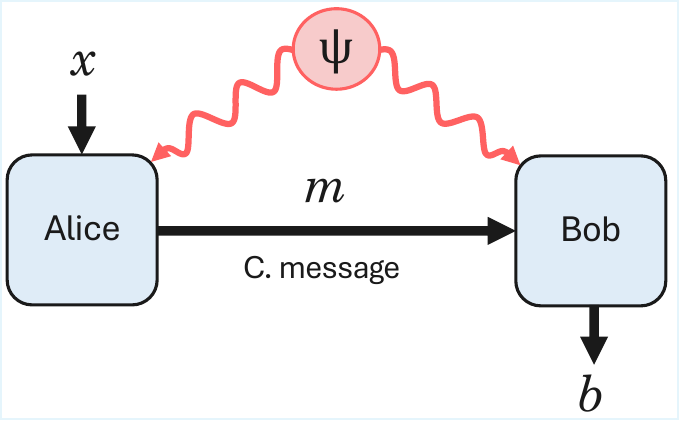}
	\caption{Prepare-and-measure scenario without receiver input assisted by an entangled state $\psi$. Alice selects input $x$, encodes it in a classical message $m$ sent to Bob, who upon receiving it, decodes the message and produce output $b$.}\label{fig_PMscenario}
\end{figure}

A quantum advantage means the violation of these inequalities. This requires the parties to share an entangled state, $\psi_{AB}$, as illustrated in Fig~\ref{fig_PMscenario}. Alice encodes her input $x$ by performing a quantum measurement, $\{A_{m|x}\}_m$, on her share of $\psi$, and then sends the measurement outcome, $m$, to Bob. Bob reads $m$ and subsequently decodes by performing a quantum measurement, $\{B_{b|m}\}_b$, on his share of  $\psi$. This leads to correlations of the form \cite{Pauwels_2022}
\begin{equation}\label{eq:EA_CC}
p(b|x) = \sum_{m=1}^{d} \Tr(A_{m|x} \otimes B_{b|m} \psi_{AB}).
\end{equation}
For these correlations to not admit a classical model, it is essential that Bob adapts his quantum measurement to $m$. To see that, note that if Bob's measurements are only classical post-processings of a single measurement, then they are by definition jointly measurable and can therefore be written $B_{b|m}=\sum_\lambda p(b\lvert m, \lambda) G_\lambda$ for some measurement $\{G_\lambda\}_\lambda$. Inserting this in \eqref{eq:EA_CC} returns a classical model.

\textit{The minimal scenario.---} 
We begin with finding the minimal PM scenario that has a  quantum advantage.  This means considering binary messages ($d=2$). Since we must have $|X|>d$, the minimal choice is three inputs for Alice. Naively, one could consider binary outcomes ($|B|=2$), but this is trivial because Alice can for each $x$ send the right answer to Bob as  $m=b$. This makes ternary outputs a natural candidate for the minimal scenario, i.e.~when the alphabets are of size $(d,|X|,|B|)=(2,3,3)$. We have used the software PANDA \cite{Lorwald_2015} to derive all the inequalities \eqref{eq:facet} that characterise the classical polytope. The only non-trivial inequality is $p(0|0)+p(1|1)+p(2|2)\leq 2$. This can be interpreted as a state discrimination task: Bob must guess Alice's input after receiving her bit. However, it is well-known that no amount of entanglement can enhance state discrimination tasks \cite{QSD_proof}, and therefore no quantum advantage is possible.

In search of the minimal scenario, we therefore add one more output for Bob, i.e.~we consider $(d,|X|,|B|)=(2,3,4)$. A full characterisation of the classical polytope reveals that there is only one new facet inequality,
\begin{equation}\label{eq:witness_fsd}
	\mathcal{S}_{1} = \frac{1}{4}\sum_{i=0}^2 \big[2p(i|i)+p(\perp|i)\big] \leq 1,
\end{equation}
where we have labeled Bob's outcomes by $b\in\{0,1,2,\perp\}$. This inequality has a very simple interpretation: if Bob is able to guess $x$ he earns two points; if he fails he earns no points; and if he instead outputs  $b=\perp$ he always earns one point regardless of $x$. Thus, it is a state discrimination task with a limited reward for discarding rounds.  

We now construct a protocol that achieves the optimal quantum advantage possible by sharing entangled qubits. Let the shared state be $\ket \psi = \sqrt{\frac{2}{3}}\ket{00}+\frac{1}{\sqrt{3}} \ket{11}$ and let $(X,Y,Z)$ be the Pauli observables. When Alice has $x\in\{ 0,1\}$, she measures the observable $\frac{1}{\sqrt{3}}((-1)^x \sqrt{2}X+Z)$ on her qubit and relays the outcome $m \in \{0,1\}$ to Bob. For  $x=2$, Alice just discards her  local state and deterministically sends the message $m=1$. Depending on the value of $m$, Bob performs one of two different measurements on his qubit: when $m=0$ he measures $X$, and when $m=1$ he measures $Z$. The outcome is either $+1$ or $-1$. Now, he constructs $b$ as follows: for $m =0$ he maps $(+1,-1)\rightarrow (0,1)$, whereas for $m=1$ he maps $(+1,-1)\rightarrow (2,\perp)$. Computing the correlations gives the violation  $\mathcal S_{1} = \frac{9+2\sqrt{3}}{12} \approx 1.0387$. We note that this decimal value was reported from a numerical search in Ref~\cite{Vieira_2023}. In our analytical construction, we see that the state is non-maximally entangled, and in fact the best protocol based on the maximally entangled states achieves at best $\mathcal S_{1}\approx 1.0295$ which is suboptimal.

However, perhaps surprisingly, larger advantages are possible by using entanglement of higher dimension.  Let Alice and Bob share the partially entangled state $\ket{\psi} = \sqrt{\frac{15}{46}}\ket{00}+ \sqrt{\frac{15}{46}}\ket{11}+\frac{2}{\sqrt{23}}\ket{22}+\frac{2}{\sqrt{23}}\ket{33}$. Alice's encoding will be associated to three dichotomic rank-2 projective measurements, whereas Bob's four possible outcomes are associated with an adaptive choice of four-dimensional basis measurements. The specific measurements are provided in the repository~\cite{code}. This protocol gives $\mathcal{S}_{1} \approx 1.0435$, which exceeds the protocol with qubit entanglement. We have shown that this is the largest value possible in quantum theory by applying semidefinite relaxations proposed in Ref~\cite{Pauwels_2022} (at level 1+AB+BB) to recover this value up to solver precision. Once again, choosing $\ket{\psi}$ as the  maximally entangled four-dimensional state is sub-optimal as it achieves only $\mathcal S_{1} \approx 1.0344$.

\textit{Certifying adaptive one-way LOCC.---} For certifying that an experiment has the capability of adaptive one-way LOCC, the above  quantum protocol is not practical because it neither uses the maximally entangled state nor has good noise-resilience. However, it turns out that both these obstacles can be  overcome by analysing quantum advantages in the next-to-minimal input/output scenario. This corresponds to choosing $(d,|X|,|B|)=(2,4,4)$ \cite{argument2}. We have characterised the classical polytope and found that two new facet inequalities appear. We focus on one of them, namely
\begin{equation}\label{eq:witness_cse}
\begin{aligned}
\mathcal S_{2} &= \frac{1}{3} \big[p(1|0) +p(2|0) +p(0|1)+p(2|1) \\
&\qquad + p(0|2)  +p(1|2) +p(3 |3)\big] \leq 1.
\end{aligned}
\end{equation}
For completeness, we discuss the other inequality and its quantum violation in Appendix~\ref{App:Sec}. The inequality \eqref{eq:witness_cse} admits a simple task-interpretation: when  Alice has $x \in \{ 0,1,2\}$, Bob wants to output one of the two symbols that are different from $x$, wheras for $x = 3$ his goal is to learn that value. 

Let us now introduce what turns out to be the optimal quantum protocol based on qubit entanglement. Its main feature is that it closely resembles the  standard protocol for violating the CHSH Bell inequality. Specifically, let Alice and Bob share the maximally entangled two-qubit state $\ket{\phi^+} = \frac{1}{\sqrt{2}}(\ket{00}+\ket{11})$. When Alice has $x= 0$ and $x = 1$, she measures her qubit with the observable $Z$ and $X$, respectively, and sends the outcome $m\in\{0,1\}$ to Bob. When $x = 2 $ and $x =3$, she instead discard her qubit and deterministically sends $m = 0$ and $m = 1$, respectively.  Bob uses $m$ to adaptively  measure his qubit with the observable  $\frac{Z+(-1)^m X}{\sqrt{2}}$, resulting in the outcome either  $+1$ or $-1$. From this, Bob constructs the final output $b$ as follows: for $m =0$ he maps $(+1,-1)\rightarrow (0,1)$, and for $m=1$ he maps $(+1,-1)\rightarrow (2,3)$. Notice that, for the inputs $x\in\{0,1\}$, the parties effectively perform the measurements that reach Tsirelson's bound for the CHSH Bell test. A direct calculation reveals that this protocol achieves the violation $\mathcal S_{2}= \frac{1}{6}\left(5+\sqrt{2}\right) \approx 1.069$. Thanks to its connection with CHSH-type nonlocality, it has significant resilience to noise.  For example, if $\phi^{+}$ is exposed to depolarising noise with visibility $v$ then a violation of Eq~\eqref{eq:witness_cse} is achieved for $v>\frac{1}{\sqrt{2}}$. If it instead is exposed to dephasing noise with visibility $v$, a violation is achieved for $v>\sqrt{2}-1$, but this is higher than for CHSH.

We emphasise that the violation of the inequality \eqref{eq:witness_cse} is not equivalent to violating the CHSH inequality. Apart from the physical scenarios being different, one also cannot map every violation of the CHSH inequality to a violation of the inequality \eqref{eq:witness_cse}. In fact, many advantages in the PM task are entirely independent of CHSH nonlocality. In particular, and in analogy with the minimal scenario, but it contrast to the CHSH inequality, the largest possible violation requires high-dimensional entanglement. By sharing a four-dimensional maximally entangled state, the parties can reach $\mathcal S_{2} \approx 1.0749$. This is achieved by Alice performing four dichotomic rank-2 measurements, and by Bob adaptively making his choice of basis measurements. The protocol is available at the repository \cite{code}. Using the semidefinite relaxations of Ref~\cite{Pauwels_2022} (at level 1+AB), we have confirmed this is the largest advantage allowed in quantum theory.

\textit{Facet inequalities.---} We conclude our analysis of the PM scenario with classical messages by providing a family of facets for the classical polytope. These are valid for any scenario associated with alphabet sizes of the form $(d,|X|,|B|)=(2,n,n+1)$, for $n\geq 3$. For $n=3$, it reduces to the inequality \eqref{eq:witness_fsd}, which we have shown to be the minimal scenario for a quantum advantage. Hence, this family of facets can be viewed as a  generalisation. They are given by 
\begin{equation}\label{facetfamily}
\begin{aligned}[t]
\mathcal S^{(n)} &= \frac{1}{2 (n-1)} \sum_{i = 0}^{n-1}\big[ (n-1) p(i|i) +p(\perp|i)\big] \leq 1,
\end{aligned}
\end{equation}
where we have written  $b\in\{0,\ldots,n-1,\perp\}$ for Bob's outcome. We observe that the task-interpretation is very similar to that of the minimal scenario; Bob is given a reward for guessing $x$ but can deterministically secure a smaller reward by outputing $b=\perp$. In Appendix~\ref{App:facets}, we first prove the bound on the right-hand-side of \eqref{facetfamily} and then prove that it is a facet for any $n$ by analysing the determinsitic strategies that saturate the bound.

\begin{figure}[t!]
	\centering
	\includegraphics[width=0.8\columnwidth]{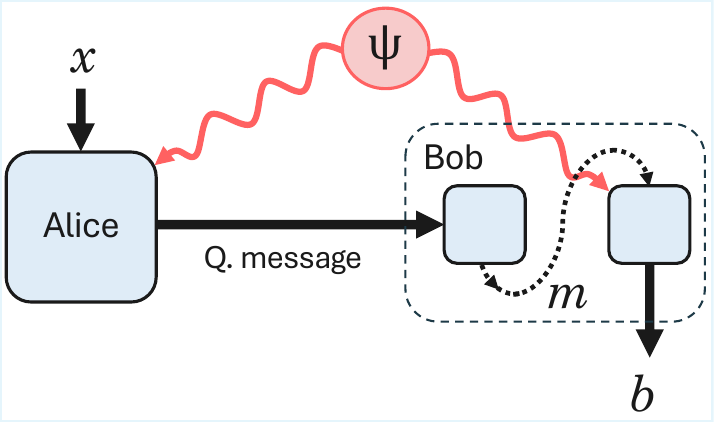}
	\caption{Prepare-and-measure scenario using quantum communication. Using the entangled state $\psi$, Alice encodes input $x$ into a quantum message sent to Bob. Upon receiving it, Bob first measures the quantum message and then use the classical read-out $m$ to select a measurement on his entangled particle.}\label{fig_QCscenario}
\end{figure}

\textit{Qubits messages and non-projective measurements.---} A natural basic question  is whether substituting Alice's bit-messages for qubit-messages leads to any interesting new features. Naturally, when qubit communication is assisted by entanglement, Alice and Bob can perform protocols in the spirit of dense coding \cite{Bennett1992, Tavakoli_2021, Pauwels_2022b} to generate much stronger correlations. However, we are going to study qubit messages in the same framework as we have studied classical messages, namely by Bob first reading out classical information from the message and then using this to advice his choice of measurement on his share of the entangled state. This is illustrated in Fig~\ref{fig_QCscenario}. Considering this situation, where Bob cannot jointly measure  his two quantum particles, is not only motivated by extending the classical communication framework into its simplest quantum setting, but also by the difficulty of actually performing general entangled measurements. Protocols that avoid the latters to enable simpler experiments have recently been implemented \cite{Piveteau2022, Piveteau2024, Bakhshinezhad2024, Zhang2025}.

In the scenario in Fig~\ref{fig_QCscenario}, Alice's encoding is a quantum channel, $\Lambda_x$, which maps her share of $\psi_{AB}$ to the qubit message that is sent to Bob. When Bob receives this qubit, he reads it by performing a measurement $\{M_m\}_m$. The classical read-out, $m$, is used to select the measurement, $\{B_{b|m}\}_b$ that Bob performs on his share of $\psi_{AB}$. Hence, the quantum  correlations take the form
\begin{equation}\label{eq:EA_QC}
	p(b|x) = \sum_{m} \Tr\left( M_{m} \otimes B_{b|m} \tau^{AB}_x \right),
\end{equation}
where $\tau^{AB}_x = (\Lambda_x \otimes \openone)[\psi_{AB}]$ is the total state given to Bob after Alice's encoding, and $\tau_x^A$ is a qubit.  

The central question is whether this can lead to correlations that cannot be generated in entanglement-assisted protocols based on classical messages. One may suspect that the answer is negative for the following reasons. Firstly, due to the Frenkel-Weiner theorem \cite{Frenkel_2015}, qubits have no advantage over bits in the PM scenario without receiver inputs, and in our model they are merely used to communicate advice about which measurement Bob should perform. Secondly, quantum correlation protocols typically use projective measurements, and whenever Bob uses such a measurement to read Alice's message, no advantage is possible: Alice could already perform the projective measurement $\{M_m\}_m$ in her lab and then simply send the bit $m$ to Bob. Hence, if qubit messages are to be useful, they must be read using qubit measurements that cannot be reduced to stochastic combinations of dichotomic measurements. This is equivalent to $\{M_m\}_m$ being a genuinely non-projective measurement \cite{Oszmaniec2017}.

We now return to the minimal scenario given by the inequality \eqref{eq:witness_fsd} and show that non-projective read-out measurements allow for quantum advantages that cannot be achieved using any protocol based on classical messages. To this end, let the parties share the partially entangled state $\ket{\psi} = \frac{3}{5} \ket{00} + \frac{4}{5} \ket{11}$. When Alice receives $x=0$, she measures the observable $Z$ on her qubit. Then, she prepares the state $\tau_0^A = \ket{1}$ and sends it to Bob. This renders $\tau_0^{AB}$ separable. For $x = 1$ and $x = 2$, Alice instead applies the Pauli unitaries $X$ and $Y$ on her share of $\psi$, respectively, and then relays the qubit to Bob. Bob measures Alice's incoming qubit with a three-outcome measurement. Specifically, he chooses the trine measurement, whose Bloch vectors are  $(\frac{\sqrt{3}}{2},0,\frac{1}{2})$, $ (-\frac{\sqrt{3}}{2},0,\frac{1}{2})$, and $(0,0,-1)$, associated with $m=0,1,2$ respectively. Based on $m$, Bob selects the measurement to be performed on his share of $\ket \psi$. For $m = 0$ and $m=1$ he measures $X$ and records either $+1$ or $-1$. He the maps this outcome to $b$ by the rule $(+1,-1) \rightarrow(1,2)$ and $(+1,-1) \rightarrow(2,1)$ for $m=0,1$ respectively. For $m = 2$ he measures $Z$ and maps $(+1,-1) \rightarrow(\perp,0)$. Evaluating this protocol gives
\begin{equation}
	\mathcal S_{1} = \frac{2}{75}(29+6\sqrt{3}) \approx 1.0505.
\end{equation}
This exceeds the limit of $\mathcal{S}_1\approx 1.0435$ which we earlier proved to be optimal for classical communication. Our numerics suggest that the above is the largest value achivable for qubit entanglement, but by using four-dimensional entanglement to assist the qubit communication, we have been able to increase the advantage even further, reaching a value of $\mathcal S_{1} \approx 1.0902$. Thus, upgrading the classical messages to quantum messages more than doubles the advantage originally achievable over classical models.

We have also analysed the inequality \eqref{eq:witness_cse} under entanglement-assisted qubit communication. The details are presented in Appendix~\ref{App:QubitS2}. Also in that scenario do we find an advantage over the best possible protocol based on entanglement-assisted classical messages. The main qualitative difference with the minimal scenario is that Bob's non-projective read-out measurement no longer corresponds to an equilateral triangle in the $XZ$-plane of the Bloch sphere, but now instead forms an isoceles triangle. Like the minimal scenario, the advantage is further boosted by using four-dimensional entanglement.

\textit{Discussion.---} We have conducted a systematic investigation of prepare-and-measure scenarios without independent receiver inputs. This is motivated by the fact that quantum advantages require   more sophisticated resources than when the receiver has inputs, namely shared entanglement and measurements adapted in real time. Our study reveals key quantum features in these scenarios. These include i) the minimal scenario for a quantum advantage, ii) the relevance of high-dimensional entanglement, and iii) the indispensable role of non-projective measurements for quantum advantages propelled by qubit messages. The last point is conceptually important beyond the confines of PM scenarios, because non-projective measurements are broadly believed to be of secondary interest in black-box studies of quantum correlations. This folklore has also been proven in some concrete scenarios, such as in steering \cite{Renner2024, Zhang2024} and qubit PM scenarios without shared entanglement \cite{Frenkel_2015, Renner2023}. Notably, one may also interpret the indispensable role of non-projective measurements as a means to certify them in semi-device-independent experiments; a topic that has received significant research attention but, unlike here, most often been based on first self-testing auxiliary states and measurements \cite{Acin2016, Mironowicz2019, Tavakoli2019b, Tavakoli2020b, Martinez2023}.

Furthermore, the scenarios in which we have characterised quantum advantages also have a distinct practical motivation since the quantum correlations serve as black-box certificates of adaptive one-way LOCC. Given the significant efforts invested in developing quantum technology based on real time classical feed-forward (see e.g. \cite{Prevedel2007, Ma2012, CarreraVazquez2024, Sakaguchi2023}), we believe that these tests will be of interest in the context of  long-distance quantum communication and benchmarking of quantum memories in networks.

\section*{End Matter}

\textit{Four-input and five-output scenario---}
For completeness we include all non-trivial facet inequalities in the third-to-minimal input/output scenario, which corresponds to choosing $(d,|X|,|B|) = (2,4,5)$. These can be found in App.~\ref{App:45facet}, toghether with the associated quantum advantage that can be reached when two-qubit entanglement is present. 

\textit{Informationally restricted correlations---}
Our discussion of quantum advantages in PM scenarios without independent receiver inputs have relevant implications for other PM scenarios, in which there is no entanglement between Alice and Bob, and where Alice's quantum messages are limited only by their entropic information content. This is known as informationally restricted quantum correlations; see Refs~\cite{Tavakoli_2020_info, Anubhav_2020, Tavakoli_2022}. Below, we first briefly introduce this framework and then connect it with the family of facet tests introduced in Eq~\eqref{facetfamily}.


Consider that Alice selects an input $x \in \{1,\dots,|X|\}$ and encodes it into a state $\rho_x$ relayed to Bob. Alice's set of preparations $\{\rho_x\}_{x=1}^{|X|}$  is limited  to carry one bit of information. While a qubit only can carry one bit of information, there are also many noisy high-dimensional quantum systems from which no more than one bit of information can be extracted. The entropic information content of Alice's states $\{\rho_x\}_x$ can be quantified by the accessible information
\begin{equation}
\mathcal{I}(X|B)= H_\text{min}(X) - H_\text{min}(X|B),  
\end{equation}
where $H_\text{min}$ is the min-entropy. The least restrictive case is when Alice's inputs are selected from a uniform prior, i.e., $p_x=\frac{1}{|X|}$. This gives $H_\text{min}(X)=\log(|X|)$. The conditional min-entropy is the largest probability with which Alice's classical label can be identified through a quantum measurement. This guessing probability is $P_g = \max_{N} \sum_x p_x \tr(\rho_x N_x)$, where $\{N_x\}$ is a measurement. Hence, for uniform prior we have $\mathcal{I}(X|B)=\log(|X|)+\log(P_g)$. It is well-known that for both classical and quantum states of dimension $d$, one has at most $P_g=\frac{d}{|X|}$, which gives $\mathcal{I}(X|B)=\log_2(d)$ bits.

A natural question is whether all correlations obtained through classical communication restricted by $\log(d)$ bits of information can be simulated by instead sending $d$-dimensional classical messages. In other words, is the polytope of the former communication scenario equivalent to that of the latter? Some facets of the polytope associated with $\log(d)$ bits informationally restricted classical correlations have been derived in \cite{Tavakoli_2022} but these are identical to known facets of the polytope of classical correlations associated with $d$-dimesional classical messages. Here, we prove that these classical correlation sets are inequivalent. This is achieved by adpting the inequalities $S^{(n)}$ in Eq~\eqref{facetfamily} to informationally-restricted correlations. By doing so, we additionally find, to our knowledge, the first tight family of inequalities for informationally-restricted communication. 


Alice randomly and uniformly selects an input $x\in\{0,\dots,n-1\}$ and encodes it in a classical preparation $\rho_x$ transmitted to Bob. Upon receiving the message, Bob performs a measurement $\{M_b\}_b$ and outputs $b\in\{0,\dots,n-1,\perp\}$. The state and measurement are taken to be diagonal in the same basis, i.e.,  $\rho_x = \sum_m p(m|x) \ketbra{m}{m}$ and $M_b = \sum_m p(b|m) \ketbra{m}{m}$. It is sufficient for Alice's variable $m$ to take the same values as Bob's output alphabet, i.e., $m \in\{0,\dots,n-1,\perp\}$. We  now seek to optimise $\mathcal S^{(n)}$ given that the set of preparations $\{ \rho_x\}_x$ is limited to carry at most one bit of information. First of all, normalisation of probability implies $p(i|i) + p(\perp|i) \leq 1$. Therefore, we have that $\mathcal S^{(n)} \leq  \frac{1}{2(n-1)}\big(n+(n-2) \max_{\{p(b|x)\}}  \sum_{i= 0}^{n-1} p(i|i) \big)$. Since we want to maximise $\mathcal S^{(n)}$ over the informationally-restricted correlations $p(i|i)$, the guessing probability reads $P_g = \frac{1}{n} \sum_m \max_x p(m|x) = \frac{1}{n} (\max_x p(\perp|x)+\sum_i p(i|i))$, where  Bob’s best guessing strategy for $m = \perp$ is to output the most probable input $x$. From the information constraint $\mathcal I  \leq 1$, we then obtain that $n P_g \leq 2$. This implies that the correlations between Alice and Bob must satisfy the following restriction $\sum_{i=0}^{n-1}p(i|i) \leq 1 + \min_i p(i|i)$, where we have optimally taken $\max_i p(\perp|i) = 1- \min_i p(i|i)$. It is optimal to consider the least restrictive case, in which $p(i|i) \equiv  p$ for all $i$. This gives $p \leq \frac{1}{n-1}$. Inserting this back into the expression for $\mathcal S^{(n)}$ yields
\begin{equation}\label{eq:info_ineq}
\mathcal S^{(n)} \leq \frac{n(2n-3)}{2(n-1)^2}.
\end{equation}
Note that the upper bound is strictly greater than one for all $n \geq 3$. Moreover, the upper bound can be reached when Alice's preparations take the following form $\rho_x = \frac{1}{n-1}\ketbra{x}{x} + \frac{n-2}{n-1}\ketbra{\perp}{\perp}$ for all $x$ and Bob measures in the same basis. This proves that (i) the inequality in Eq~\eqref{eq:info_ineq} is a family of tight inequalitites, and (ii) the classical polytope associated with dimension-restricted communication is strictly smaller than the corresponding one associated with informationally-restricted communication. 

\begin{acknowledgments}
We thank Emmanuel Zambrini Cruzeiro and Amro Abou-Hachem for guidance on how to use the software PANDA. We thank Anubhav Chaturvedi and Martin J.~Renner for discussion. This work is supported by the Knut and Alice Wallenberg Foundation through the Wallenberg Center for Quantum Technology (WACQT), the Swedish Research Council under Contract
No. 2023-03498 and the Swedish Foundation for Strategic Research.
\end{acknowledgments}

\bibliography{references_minimalEACC}

\begin{thebibliography}{45}
\bibitem [{arg()}]{argument2}%
  \BibitemOpen
  \href@noop {} {\bibinfo {title} {Note that the $(d,|{X}|,|{B}|)=(2,4,3)$
  scenario is uninteresting because classical polytope can effectively be
  reduced to that in the scenario $(d,|{X}|,|{B}|)=(2,3,3)$}}\BibitemShut
  {NoStop}%
\end{thebibliography}

\appendix

\section{The second facet in the 4-input-4-output scenario}\label{App:Sec}
We introduce the second facet associated the $(d,|X|,|B|)=(2,4,4)$ scenario. Alice encodes her inputs $x \in \{0,1,2,3\}$ into a single bit sent to Bob, who upon receiving the message decodes it and outputs $b\in\{0,1,2,3\}$. The second non-trivial facet of this scenario reads
\begin{equation}
\begin{aligned}
&\mathcal S_{3} = \frac{1}{4} \big[2 p(0|0) +p(2|0) +2p(1|1) + p(2|1) +p(2|2)  \\
&\qquad \quad +p(3|2)+p(3|3)\big] \leq 1.
\end{aligned}
\end{equation}
The upper bound is valid for all classical models. The optimal two-qubit state is constructed as follows $|\tilde \psi\rangle = U_1 \otimes U_2 \ket{\psi}$. The unitaries $U_1,U_2$ and Schmidt decomposition $\ket \psi$ reads
\begin{equation}
\begin{aligned}
&U_1 \approx (-0.418 + 0.478i)Z +  (-0.509 + 0.581i)X\\
&U_2 \approx (-0.555 + 0.634i)\openone  -(0.405 + 0.355i)Y\\
&\ket \psi \approx 0.7737 \ket{00} + 0.6335 \ket{11}.
\end{aligned}
\end{equation}

When Alice receives $x = 0,2$ she measures projectively. The observables assosiated with the measurement is $0.998 Z -0.06 X$ when $x=0$, and $-(0.237 Z +0.972 X)$ when $x=2$. After measuring Alice sends the resulting measurement outcome $m \in \{0,1\}$ to Bob. In contrast, for $x = 1$ and $x = 3$ she discards her state and deterministically sends Bob the message $m = 0$ and $m = 1$, respectively. When Bob receives message $m = 0$, he measures the observable $X$ and associates outcome $(+1,-1) \rightarrow (2,1)$, whereas for $m =1$ Bob approximately measures the observable $0.984 Z+0.1784 X$, and maps the outcomes $(+1,-1) \rightarrow (3,0)$. The protocol based on qubit entanglement can at most reach
\begin{equation}
\mathcal S_{3} \approx 1.0446
\end{equation}

However, by using ququart entanglement we can increase the value of the correlation inequality to $\mathcal S_{3} \approx 1.04771$. This is achieved by Alice and Bob sharing a partially entangled state with Schmidt decomposition
\begin{equation}
\ket{\psi} = 0.5502 \ket{00} + 0.5502\ket{11} + 0.4442 \ket{22} +0.4442 \ket{33}.
\end{equation}
In this case, Alice performs four dichotomic rank-two measurements, while Bob performs two four-dimensional basis measurements. The protocol is available at the repository \cite{code}. Using the semidefinite relaxations of Ref~\cite{Pauwels_2022} (at level 2), we obtain an upper bound that match our explicit construction up to four decimals.

\subsection{Quantum communication}
By using a qubit message, we can enhance the adaptive communication advantage. Let Alice and Bob share the maximally entangled state $\ket{\phi^+}$. Alice applies unitaries $X$ and $Y$ to her qubit when $x = 0$ and $x = 1$, respectively, and send the resulting state to Bob.  When $x = 2$ and $x = 3$ Alice measures her local state with observable $Z$. In the former case, she then sends $\frac{1}{\sqrt{2}}(\ket{0}+\ket{1})$ to Bob. In the latter case, her qubit message depends on her outcome $(+1,-1)$. When she receives $+1$ she sends $\frac{1}{\sqrt{10}}(3\ket{0}+\ket{1})$, whereas for $-1$ she sends $\frac{1}{\sqrt{10}}(\ket{0}+3\ket{1})$. When Bob receives Alice's qubit he perform a three-outcome POVM $\{M_m\}_m$ on the incoming state according to
\begin{equation}
\begin{aligned}
&M_0 =  \frac{3}{8}(\openone-X)\\
&M_1=  \frac{1}{16}(5\openone +3X-4Z)\\
&M_2 =  \frac{1}{16}(5\openone+3X+4 Z)
\end{aligned}
\end{equation}
Given the read-out $m = 0$, Bob then measures $X$ on his share of $\ket{\phi^+}$ and associates the measurement outcome to his final output values as follows $(+1,-1) \rightarrow(1,0)$. When $m = 1,2$, he measures $Z$ on his particle and outputs  $(+1,-1) \rightarrow(2,3)$ and $(+1,-1) \rightarrow(3,2)$ when $m=1,2$. This results in that
\begin{equation}
\mathcal S_{\text{3}} = \frac{17}{16}.
\end{equation}
Finally, if we allow for ququart entanglement the communication advantage can be improved to $\mathcal S_{3} \approx 1.0945$

\section{A family of facet inequalities}\label{App:facets}
We derive a family of facet inequalities. The family is parameterised by an integer $n \geq 3$. Consider that Alice is given $n$ inputs and Bob is given $n+1$ outputs. We denote Alice's inputs by $x\in \{0,\dots,n-1\}$ and Bob's outputs by $b\in\{0,\dots,n-1,\perp\}$. Alice is limited to encode her input $x$ into a binary message $m\in\{0,1\}$ sent to Bob, who then decodes it and produces an output $b$. In this setting, the following correlation function is a facet of the local polytope 
\begin{equation}
\begin{aligned}[t]
\mathcal S^{(n)} &= \frac{1}{2 (n-1)} \sum_{i = 0}^{n-1} \big[ (n-1) p(i|i) +p(\perp|i)\big] \leq 1.
\end{aligned}
\end{equation}
Here, $p(b|x)$ is the probability that Bob outputs $b$ given Alice's input $x$. We first discuss the main features of classical correlations. Then we prove that the upper bound for this inequality is $\beta = 1$ for all $n$, and that it is satisfied by all classical deterministic strategies. Thereafter, we provide an explicit procedure to construct a set of linearly independent strategies that saturates $\mathcal S^{(n)} = 1$, and then use this set to prove that $\mathcal S^{(n)}$ is a facet of the local polytope.

\subsection{Classical correlations}
The correlations of the adaptive prepare-and-measure-scenario admits a classical model if they can be decomposed on the form $p(b|x) = \sum_\lambda p (\lambda) p_\lambda(b|x)$, where $p(\lambda)$ is a probability distribution and 
\begin{equation}
p_\lambda(b|x) = \sum_{m=0,1}  p(m|x,\lambda) p(b|m,\lambda).
\end{equation}
Here, $ p(m|x,\lambda) $ and $p(b|m,\lambda)$ are known as response functions. In general, these are convex combinations of deterministic response functions. However, by absorbing all randomness into $p(\lambda)$, we can w.l.g. focus on the deterministic response functions \cite{Tavakoli_2024}. We emphasize this by writing
\begin{equation}\label{eq:det}
\begin{aligned}
p_\lambda(b|x) &= \sum_{m=0,1}  D(m|x,\lambda) D(b|m,\lambda) \\
& =\sum_{m=0,1}  \delta(f_\lambda(x) = m) \delta(g_\lambda(m) = b).
\end{aligned}
\end{equation}
The functions $f_\lambda$ and $g_\lambda$ deterministically assign the outcomes of Alice and Bob based on their respective inputs and the shared variable $\lambda$. Note that $p_\lambda(b|x) \in \{0,1\}$ such that $\sum_b p_\lambda(b|x) = 1$ for all $x, \lambda$. 


Geometrically, the set of probabilities that admit a classical model is a polytope $\mathcal L$ of dimension $\text{dim}(\mathcal L) = n^2$. The polytope only depends on the number of measurement settings and measurement outputs but not on the way we label them \cite{Pironio_2014}. Moreover, the local polytope is fully characterised by its vertices. The vertices are points in the classical probability space and the probability distribution associated with each determinstic strategy $\lambda$ corresponds to a vertex of the polytope
\begin{equation}\label{eq:vertex}
\vec p_\lambda = \{p_\lambda(b|x)\}_{b,x}.
\end{equation}

\subsection{The classical bound}
We show that $\mathcal S^{(n)}$ cannot exceed unit for all classical correlations. Using that $\mathcal S^{(n)}$ is linear in the correlations, the functional reaches its maximum value at the vertices of the polytope $\mathcal L$. This implies that
\begin{equation}
\begin{aligned}[t]
\mathcal S^{(n)} &= \frac{1}{2 (n-1)}\sum_\lambda p(\lambda) \sum_{i = 0}^{n-1}  \big[ (n-1) p_\lambda(i|i) +p_\lambda(\perp|i)\big]  \\
&\leq \max_\lambda  \frac{1}{2 (n-1)} \sum_{i = 0}^{n-1} \big[ (n-1) p_\lambda(i|i) +p_\lambda(\perp|i)\big].
\end{aligned}
\end{equation}

Since Alice is limited to communicate only one bit of information, Bob can at most distinguish two of Alice's inputs. This results in that we only have to study three cases. 

The first relevant case is when Bob construct his final output $b$ from the binary message as follows: $(0,1)\rightarrow(j,k)$, where $j,k \in \{0,...,n-1\}$. In this case, $p(\perp|x) = 0$ for all $x$. It follows that the correlation functional reduces to the standard quantum state discrimination task
\begin{equation}
\mathcal S^{(n)} = \frac{1}{2} \sum_{i = 0}^{n-1} p(i|i).
\end{equation}
Since Bob only can distinguish two of Alice's inputs, this directly leads to the classical limit $\mathcal S^{(n)} \leq 1$. There is $\binom{n}{2}2^{(n-2)}$ deterministic strategies for which the parties can achieve the classical limit.

The second relevant case is when Bob constructs his output as $(0,1)\rightarrow(j,\perp)$ and $j \in \{0,...,n-1\}$.  In this case, Bob only distinguish one of Alice's inputs. Hence, the correlation function simplifies to
\begin{equation}
\mathcal S^{(n)} = \frac{1}{2} \delta(0=f(j)) + \frac{1}{2(n-1)}\sum_{i = 0}^{n-1} \delta(1=f(i)) \leq 1. 
\end{equation}
For each distinct choice of $j$, there is one deterministic strategy $f(x)$ that saturates the classical bound, namely when $f(j) = 0$ and $f(i) = 1$ for all $i \neq j$.

In the last case, Bob constructs his output from the binary message as follows: $(0,1)\rightarrow(\perp,\perp)$. This results in that the correlation function simplifies to
\begin{equation}
\mathcal S^{(n)} = \frac{1}{2(n-1)} \sum_{i = 0}^{n-1} p(\perp|i) \leq \frac{n}{2(n-1)} < 1.
\end{equation}

Thus, $\mathcal S^{(n)}\leq 1$ for all possible determinstic strategies. Moreover, we find that the total number of deterministic strategies that saturates $\mathcal S^{(n)} = 1$ is given by
\begin{equation}
N = \binom{n}{2}2^{(n-2)} + n.
\end{equation}
Note that each of the $N\geq n^2$ strategies corresponds to a vertex in the local polytope. However, since the dimension of the local polytope is $\text{dim}(\mathcal L) = n^2$, there exists at most $n^2$ linearly independent vertices among $N$.

\subsection{Linearly independent strategies}
Through an explicit construction, we now show that there exists $n^2$ linearly independent vertices for which $\mathcal S^{(n)}= 1$. We denote the set of these strategies by $\tilde P$.

As already mentioned above, there is $n$ different ways to play the second strategy in which Bob only discriminates one of Alice's inputs and outputs "inconclusive" in the remaining cases. For each value $j \in \{0,\dots,n-1\}$, the classical correlations read
\begin{equation}
\begin{aligned}
p(b|x) = \delta(x=j)\delta(b=j)+ \delta(x\neq j)\delta(b = \perp).
\end{aligned}
\end{equation}
Since none of the associated vertices can be written as a linear combination of the others, each of them is in the set $\tilde P$.

Next, we study the second case in which the parties want to perform state discrimination. First, consider that Bob aim to distinguish Alice's inputs $x \in\{0,k\}$ for some $k \in\{1,\dots,n-1\}$. When $x = 0(x=k)$, Alice optimally sends $m = 0(m=1)$ to Bob, who then outputs $b = 0 (b=k)$. While this protocol fully fixes Bob's deterministic response function to $g(0) = 0$ and $g(1) = k$, it does not specify Alice's message $m$ when $x \notin \{0,k\}$. The set of messages compatible with the described strategy is
\begin{equation}\label{eq:mess}
\begin{aligned}
\vec m_\lambda =  [0,f_\lambda(1),...,f_\lambda(k-1),1,f_\lambda(k+1),...,f_\lambda(n-1)].
\end{aligned}
\end{equation}
However, we are only interested in the subset of strings $\{\vec m_\lambda\}$ that are linearly independent. These are obtained by restricting ourselves to response functions on the form
\begin{equation}\label{eq:det_strat}
\begin{aligned}
&f_\lambda(x) =  \delta(x = \lambda), \qquad \quad 0<x<k \\
&f_\lambda(x) = \delta(x-1 = \lambda), \qquad x>k,
\end{aligned}
\end{equation}
where $\lambda \in \{0,...,n-2\}$. For this choice, the local correlations read
\begin{equation}
\begin{aligned}
&x = 0: \qquad \quad p_\lambda(0|0)= 1\\[0.5em] 
&x= k: \qquad \quad p_\lambda(k|k) = 1 \\[0.5em] 
&0<x < k: \quad p_\lambda(b|x) = \delta(x \neq \lambda) \delta(b =0)  \\
&\qquad \qquad \qquad \qquad \quad \ + \delta(x = \lambda) \delta(b =k) \\[0.5em] 
&x> k:  \qquad \quad p_\lambda(b|x) = \delta(x-1 \neq \lambda) \delta(b =0)  \\
&\qquad \qquad \qquad \qquad \quad  \ \ + \delta(x-1 = \lambda) \delta(b =k).
\end{aligned}
\end{equation}
That is, Bob generate the output
\begin{equation}
\begin{aligned}
&b = 0 \quad \text{when } 
\begin{cases}
&x \neq \lambda \qquad \ \ \text{if} \ x <k \\ 
&x-1 \neq \lambda \quad \text{if} \ x >k 
\end{cases}\\
&b = k \quad \text{otherwise}.
\end{aligned}
\end{equation}
Representing the associated vertex $\vec p_\lambda = \{p_\lambda(b|x)\}_{b,x}$ as a table, where the rows represent Alice's input $x$ and the columns represent Bob's output $b$, we have that
\begin{equation}\label{eq:vertices}
\begin{matrix}
\vec p_\lambda = [&1 & \cdots & 0 & 0 & 0 & \cdots & 0; \\
& \delta(1 \neq \lambda)& \cdots & 0 &  \delta(1 = \lambda) & 0 & \cdots & 0; \\
&\vdots & & \vdots & \vdots & \vdots & & \vdots \\
&\delta(k-1 \neq \lambda) & \cdots & 0 &\delta(k-1 = \lambda)& 0 & \cdots & 0; \\
&0 & \cdots & 0 & 1 & 0 & \cdots & 0; \\
&\delta(k \neq \lambda) & \cdots & 0 &\delta(k = \lambda) & 0 & \cdots & 0; \\
&\vdots & & \vdots & \vdots & \vdots & & \vdots \\
&\delta(n-2 \neq \lambda) & \cdots & 0 & \delta(n-2 = \lambda)& 0 & \cdots & 0].
\end{matrix}
\end{equation}
Note that only the zeroth and $k$:th column contains non-zero elements. Following this strategy, we find that for each choice of $k \in\{1,\dots,n-1\}$ there is $(n-1)$ linearly independent classical strategies that yields $\mathcal S^{(n)} = \frac{1}{2}(p(0|0) + p(k|k)) = 1$. Hence, in total this scheme gives $(n-1)^2$ additional vertices to the set $\tilde P$.

So far, the parties has always used the message $m = 0$ to acheive $p(b=0|0) = 1$. We find $n-1$ more linearly independent vertices by considering the case in which Bob fails to output the correct value $b = 0$ given $x = 0$. First, consider that Bob instead correctly distinguish Alice's inputs $x = 1,2$, by the parties playing the following strategy: $f(x) = \delta(x \in \{0,1\})$, and $g(0) = 2$ and $g(1) = 1$. This yields that $\mathcal S^{(n)} = \frac{1}{2}(p(1|1)+p(2|2)) =1$ and $p(b=1|0) = 1$. Next, consider that Bob correctly distinguish $x = 1,k$, with $k \in \{2,...,n-1\}$. In this case the parties perform the following strategy: $f(x) = \delta(x \in \{ 0,k\})$ and $g(0) = 1$ and $g(1) = k$, with $k \in \{2,...,n-1\}$. This yields that $\mathcal S^{(n)} = \frac{1}{2}(p(1|1)+p(k|k)) = 1$ and $p(k|0) = 1$ for all $k>1$.

At this point, we have found $ (n-1)^2+(n-1)+n = n^2$ linearly independent vertices that saturates the local bound $\mathcal S^{(n)} = 1$. As the dimension of the local polytope is $n^2$, there exists no more. We denote these vertices by $x_i$, and thus take the set $\tilde P$ to be $\tilde P = \{x_i\}_{i=1}^{n^2}$.

\subsection{Proof of facet}
We now show that the set of vertices in $\tilde P$ spans a facet. Since the collection of points in $\tilde P$ saturates the inequality $\mathcal S^{(n)} \leq 1$, these points lies in a face of $\mathcal L$ \cite{Ziegler}.  The dimension of the face $\tilde P$ is the dimension of its affine hull, i.e., $\dim(\text{aff}(\tilde P))$, where
\begin{equation}
\text{aff}(\tilde P) = \{  \sum_{i=1}^{n^2} \lambda_i x_i \ | \sum_{i=1}^{n^2} \lambda_i = 1\},
\end{equation}
for some coefficients $\lambda_i$. Specifically, every point $x$ in the affine hull of $\tilde P$ can be written as $ x = x_1 + \sum_{i=2}^{n^2} \lambda_i (x_i-x_1)$. Hence, we have that~\cite{Ziegler}
\begin{equation}
\text{dim}\big(\text{aff}(P)\big) = \text{dim}\big( \text{span}\{x_2-x_1,...,x_{n^2}-x_1\}\big)
\end{equation}
However, since the collection of vertices $\{x_i\}_i$ are linearly independent, this gives 
\begin{equation}
\text{dim}(\tilde P) = \text{dim}(\mathcal L)-1=n^2-1.
\end{equation}
This proofs that the equality $\mathcal S^{(n)} = 1$ is a facet.

\section{Violation of {$\mathcal S_2$} using quantum communication}\label{App:QubitS2}
We show that entanglement-assisted qubit communication can lead to higher violation of $\mathcal S_2$ than possible with classical communication. Let Alice and Bob share the maximally entangled state $\ket{\phi^+}$. When $x = 0$, Alice measures $Z$ on her share of $\ket{\phi^+}$ and then sends the state $\ket{1}$ to Bob. When $x = 3$, she again measures $Z$ on her qubit. Depending on her measurement output $(+1,-1)$, she now sends one of two states to Bob. For $+1$ she relays the state $\ket{0}$, whereas for $-1$ she relays $\openone/2$. In contrast, when Alice receives inputs $x = 1 $ and $x =2$ she applies the unitaries $X$ and $Y$ to her local share of $\ket{\phi^+}$ and then sends it to Bob. Hence, for $x=0,3$ Alice renders the share state separable, whereas for $x=1,2$ it remains maximally entangled. Bob measures the incoming qubit message in three directions corresponding to an isosceles triangle in the XZ-plane. Specifically, the measurement associated with measurement outcome $m=0$ reads $\frac{3}{8}(\openone+Z)$ whereas those associated with $m=1,2$ reads $\frac{1}{16}(5\openone + ((-1)^m 4X-3Z))$. Given the read-out $m=0$, Bob then measures his share of $\ket{\phi^+}$ with the observable $Z$, and associates the measurement outcomes $+1$ and $-1$ to his final output as follows: $(+1,-1) \rightarrow (3,0)$. For $m = 1,2$ Bob instead measures the observable $X$ and maps $(+1,-1)\rightarrow (1,2)$ when $m = 1$ and $(+1,-1)\rightarrow (2,1)$ when $m=2$. This gives
\begin{equation}
	\mathcal S_{\text{2}} = \frac{13}{12} \approx 1.0833.
\end{equation}
By using shared ququart maximally entanglement we can further reach a value of $\mathcal S_{\text{2}} = 1.1320$.
	
\section{Facet inequalities for 4-input-5-output scenario}\label{App:45facet}
For completeness we provide a full list over the non-trivial facet inequalities found in the $(d,|X|,|B|)=(2,4,5)$ scenario, see table \ref{tab:45scenario}. These are found using the software PANDA \cite{Lorwald_2015}. Here, $p(b|x)$ is the probability that Bob outputs $b$ given Alice's input $x$. We also provide the optimal quantum violation that was found when the parties share two-qubit entanglement and Alice is limited to transmit one classical bit of information to Bob. The quantum violation is found be performing an alternative convex search based on semidefinite programming routines (see e.g. review \cite{Tavakoli_2024}). Specifically, the search optimise the inequalities over all two-qubit entangled states and all local measurements. The routine is decomposed into three sub-routines: (i) optimisation over Alice's encoding measurements $\{A_{m|x}\}$, (ii) optimisation over Bob's decoding measurements $\{B_{b|m}\}$, and (iii) optimisation over the shared state $\psi_{AB}$. The three routines are iterated until convergence is reached.

\onecolumngrid

\begin{table}[ht!]
\begin{tabular}{|l|l|l|}
\hline
    & Facet inequalities  &Quantum violation \\ \hline
1.  &     $ 2p(1|0) + p(2|0) + p(3|0) + p(2|1) + p(3|1) + 2p(4|1) - p(1|2) - p(3|2) - p(4|2) - p(1|3) - p(2|3) - p(4|3)   \leq 2    $   &         2.2071       \\ \hline
2. & $    2p(1|0) + 2p(2|0) + p(3|0) + p(3|1) + 2p(4|1) - 2p(1|3) - 2p(2|3) - p(3|3) - 2p(4|3)         \leq 2     $ &      2.1547          \\ \hline
3. &   $  2p(1|0) + p(2|0) + p(3|0) + p(2|1) + p(3|1) + 2p(4|1) - 2p(1|3) - p(2|3) - p(3|3) - 2p(4|3)       \leq 2$     &          2.1547      \\ \hline
4. &      $2p(1|0) + 2p(2|0) + p(3|0) + p(3|1) + 2p(4|1) - p(1|2) - p(2|2) - p(4|2) - p(1|3) - p(2|3) - p(3|3) - p(4|3)     \leq 2   $  &     2.1784           \\ \hline
5. &        $2p(1|0) + p(2|0) + p(3|0) + p(2|1) + 2p(4|1) + p(3|2) - 2p(1|3) - p(2|3) - 2p(3|3) - 2p(4|3)  \leq 2     $  &    2.1784            \\ \hline
6. &         $2p(1|0) + p(2|0) + p(3|0) + p(2|1) + p(3|1) + 2p(4|1) - p(1|2) - p(4|2) - p(1|3) - p(2|3) - p(3|3) - p(4|3)      \leq 2  $   &          2.1784      \\ \hline
7. &      $ p(1|0) + p(2|0) + p(3|0) + p(1|1) + p(2|1) + p(4|1) + p(3|2) + p(4|2) - p(1|3) - p(2|3) - p(3|3) - p(4|3)     \leq 2 $      &        2.2071         \\ \hline
8. &     $4p(1|0) + 2p(2|0) + p(3|0) + 2p(2|1) + p(3|1) + 4p(4|1) + 2p(3|2) - 4p(1|3) - 2p(2|3) - 3p(3|3) - 4p(4|3)   \leq 4  $ &       4.3094        \\ \hline
9. &      $  2p(1|0) + p(2|0) + p(2|1) + 2p(3|1) - p(1|2) - p(3|2) - p(1|3) - p(2|3) - p(3|3)     \leq 2    $   &         2.1784          \\ \hline
10. &       $p(1|0) + p(2|0) + p(1|1) + p(3|1) + p(2|2) + p(3|2) - p(1|3) - p(2|3) - p(3|3)  \leq 2   $  &     2.2071           \\ \hline
11. &   $ 3p(1|0) + p(2|0) + p(2|1) + 3p(3|1) + p(2|2) + 3p(4|2) - 3p(1|3) - 2p(2|3) - 3p(3|3) - 3p(4|3)     \leq 3    $   &     3.1432           \\ \hline
12. &    $ 2p(1|0) + 2p(2|0) + p(3|0) + p(1|1) + p(3|1) + 2p(4|1) + p(2|2) + p(3|2)+ $ & \\
&$ +p(4|2) - 2p(1|3) - 2p(2|3) - p(3|3) - 2p(4|3)     \leq 3    $ &      3.1784         \\ \hline
13. &      $     2p(1|0) + p(2|0) + p(3|0) + p(1|1) + p(2|1) + 2p(4|1) + p(2|2) + 2p(3|2) +$ & \\
&$ p(4|2) - 2p(1|3) - p(2|3) - 2p(3|3) - 2p(4|3)   \leq 3 $ &        3.1784       \\ \hline
14. &     $ 4p(1|0) + 2p(2|0) + p(3|0) + 2p(2|1) + p(3|1) + 4p(4|1) - p(1|2) + 2p(3|2) - $  &
\\&$p(4|2) - 3p(1|3) - 2p(2|3) - 3p(3|3) - 3p(4|3)    \leq 4    $ &      4.3202         \\ \hline
15. &     $  3p(1|0) + 3p(2|0) + 2p(3|0) + 3p(1|1) + 2p(3|1) + 3p(4|1) + 3p(2|2) +$ &\\
       & $ 2p(3|2) + 3p(4|2) - 3p(1|3) - 3p(2|3) - 2p(3|3) - 3p(4|3)   \leq 6   $  &      6.6213           \\ \hline
16. &    $ 3p(1|0) + 2p(2|0) + p(3|0) + p(1|1) + 2p(3|1) + p(4|1) + 2p(2|2) + p(3|2) + $ & \\
&$3p(4|2) - 3p(1|3) - 2p(2|3) - 2p(3|3) - 3p(4|3)         \leq 4 $  &    4.3371            \\ \hline
17. &   $ 4p(1|0) + 3p(2|0) + 2p(3|0) + 2p(1|1) + 2p(3|1) + 2p(4|1) + 3p(2|2) + 2p(3|2) + $ &\\
&$ 4p(4|2) - 4p(1|3) - 3p(2|3) - 2p(3|3) - 4p(4|3)          \leq 6$  &        6.5298         \\ \hline
18. &     $   4p(1|0) + 3p(2|0) + 2p(3|0) + 2p(1|1) + 2p(3|1) + 3p(4|1) + 3p(2|2) + 2p(3|2) + $ &\\
&$ 3p(4|2) - 4p(1|3) - 3p(2|3) - 2p(3|3) - 4p(4|3)      \leq 6 $ &      6.5733         \\ \hline
\end{tabular}
\caption{List over non-trivial facet inequalitites in the prepare-and-measure scenario in which Alice's receives $x\in \{0,1,2,3\}$ and Bob produce outputs $b \in\{0,1,2,3,4\}$.}
\label{tab:45scenario}
\end{table}

\end{document}